\documentclass[RNAAS]{aastex62}

\begin{document}

\title{How cost impacts equitable participation in astronomy outreach events} 

\correspondingauthor{Melanie Archipley}
\email{melanie@iayc.org}

\author[0000-0002-0517-9842]{Melanie Archipley}
\altaffiliation{The International Workshop for Astronomy e.V., T{\"u}rkenstra{\ss}e 12, Garching, Munich, 85748, DE}
\affiliation{Dept. of Astronomy, University of Illinois Urbana-Champaign, \\ 
1002 West Green Street, Urbana, IL, 61801, USA}

\author[0000-0002-8970-3065]{Hannah S. Dalgleish}
\altaffiliation{The International Workshop for Astronomy e.V., T{\"u}rkenstra{\ss}e 12, Garching, Munich, 85748, DE}
\affiliation{Dept. of Physics, University of Namibia, \\ Pionierspark, Windhoek, Namibia}

\begin{abstract}
    The International Astronomical Youth Camp (IAYC) is an astronomy education outreach event with more than 50 years of history and over 1,700 unique participants from 81 nationalities. The International Workshop for Astronomy e.V. (IWA) is the non-profit organization behind the IAYC, established in 1979 and based in Germany. The IAYC's unprecedented longevity in a rapidly globalizing world has meant that financial inequities decreases the reach of the camp to people from the Global South compared to Global North countries. Though nationalities represented per camp has increased steadily since its inception, the share of participants from eastern Europe and Africa has dropped, while those from western Europe and North America have increased. This note examines how camp cost, location, and leadership affects nationality diversity amongst participants, and how astronomy outreach events must reckon with funding for less privileged participants with limited access to resources. 
\end{abstract}

\keywords{astronomy education}

\section{Introduction}
The International Astronomical Youth Camp (IAYC) is a three-week summer camp where participants aged 16-24 pursue independent inquiry on astronomical topics. There are no required lectures or exams, and participants report their findings in \LaTeX\ at the end of the camp. The application consists of a letter of motivation --- resumes, transcripts, or letters of recommendation are not requested --- in order to welcome people with little prior astronomy experience.

The first IAYC took place in 1969 in West Germany, with 40 participants from 5 unique nationalities (27 participants were German). After a decade, The International Workshop for Astronomy e.V. (IWA) was founded as the body behind the camp. As written in the statues of IWA, its mission is as follows:

\begin{enumerate}
    \item \textit{To spread astronomical knowledge and to teach young people to work scientifically on their own.}
    \item \textit{Promotion of international collaboration and agreement, particularly for astronomical youth work.}
\end{enumerate}

In the past five decades, there have been 55 camps with at least one per year until 2020 (cancelled due to the COVID-19 pandemic). The 2019 camp had 65 participants from 26 different nationalities with the largest representation being 15 Spanish participants. The increase in nationalities per camp is a success with regards to IWA's second statue, with about five nationalities added every 10 years. However, the ethnic makeup of the camp is not diverse (84.7\% of IAYC survey respondents\footnote{The IAYC survey was completed by 307 former participants between 16 August 2017 and 21 September 2019. Many of the survey’s findings are published in \cite{2018capc.conf..206D,2019NatAs...3.1043D} and \citeauthor{archipley2021}, in press.} were white), and IWA is exploring how the organization could improve engagement from participants from disadvantaged areas of the world, such as Africa, eastern Europe, Asia, and South America. We use gross national income per capita (GNI) to examine the nationalities of participants, host country, and camp cost over the IAYC's long history. 

\section{Participant nationalities and GNI}
GNI reflects the average annual pretax income of a country's citizens. It is a cursory indicator for the average standard of living and is linked to other measures of a country's economic and social health. For our analysis, we use a cutoff of 30,000 USD in 2019 for ``low'' and ``high'' GNI as reported by the \cite{WorldBank} (to give an indication, Spain's 2019 GNI is 30,390 USD and Cyprus' 2019 GNI is 27,710 USD). For former countries not listed in 2019, such as Deutsche Demokratische Republik, Yugoslavia, and Czech and Slovak Federative Republic, we consider them in the ``low GNI'' group.

\begin{figure}[h!]
    \begin{center}
        \includegraphics[width=1\textwidth]{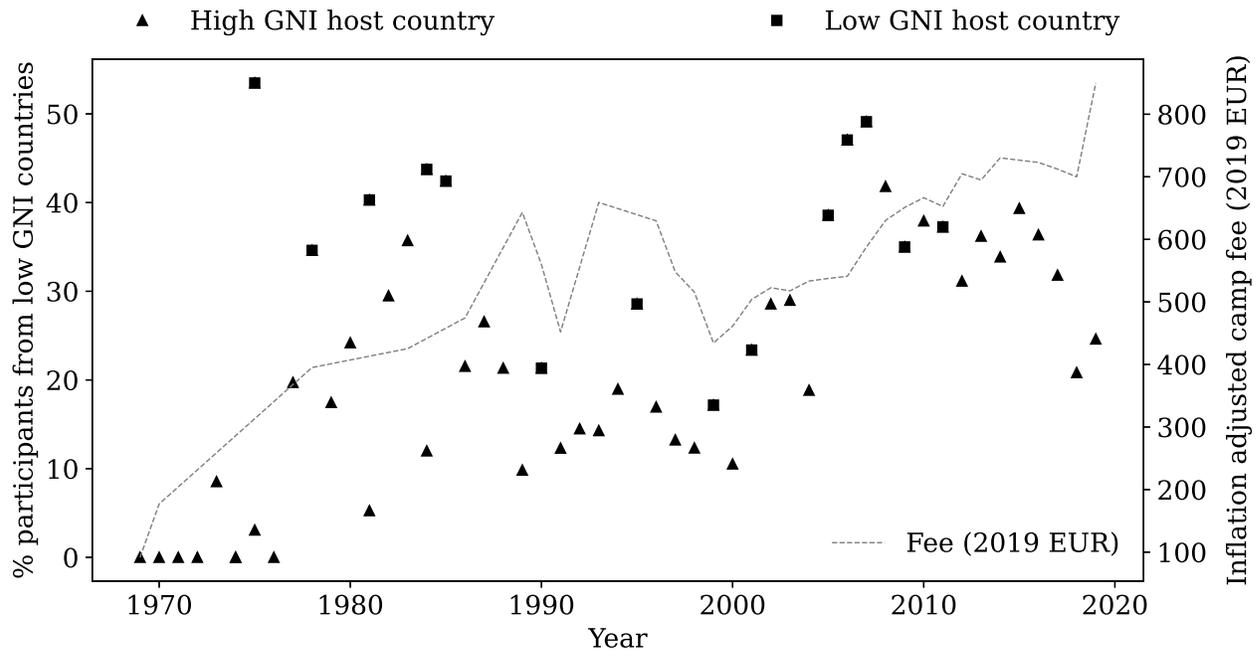}
        \caption{For each year the IAYC has taken place, we examine the fraction of participants from low GNI countries (defined as $<30,000$ USD in 2019) (\textit{black points}). We also indicate which years the camp was hosted in high and low GNI countries by triangles and squares, respectively. The inflation adjusted camp fee is shown for years we have data (\textit{gray dashed line}).  \label{fig:lowGNI_cost}}
    \end{center}
\end{figure}

Figure \ref{fig:lowGNI_cost} shows the percentage of IAYC participants from low GNI countries and the inflation adjusted camp fee per year. The majority of participants have come from high GNI countries in 54 out of 55 camps (the one exception, in 1975, was held in Tunisia). Between 2005 and 2011, the IAYC saw proportionally more participants from low GNI countries compared to recent years. We also note that for the seven years where at least 40\% of participants were from low GNI countries, six were hosted in low GNI countries (Czech Republic, Yugoslavia, Egypt, and Tunisia, with repeats). The last time the camp was hosted in a low GNI country was in 2011, and the proportion of participants from these countries has been on a downward trend since 2007. We infer that people from low GNI countries are able to attend more often when the camp is hosted in a low GNI country. This could be for several reasons, such as reduced operating costs (and therefore lower camp fee), as well as reduced travel costs for participants in local and neighboring low GNI countries.

In 2019, the camp fee was raised to 850 EUR after it remained at 690 EUR (not inflation adjusted) for the preceding five years. The IAYC has particular needs resulting in expensive camp houses: they must have beds for $\sim75$ people for three weeks, provide three daily meals at odd hours, and be located in rural areas to avoid light pollution. Although camp houses located in low GNI countries are less expensive, IWA needs general coordinators who can speak the local language, which limits the number of countries where the IAYC can be organized. Furthermore, the makeup of the IAYC leader team is limited to previous participants, so the organization risks trapping itself in a loop where the camp fee is raised to host the IAYC in a high GNI country, thus limiting potential participants from low GNI countries, meaning leadership remains exclusive to people with high GNI nationalities, further reducing chances of being able to return to a low GNI host country. 

\section{Looking forward}
In addition to operating concerns that motivate the need to attract people from low GNI countries, it is imperative to make astronomy outreach accessible. Astronomy education may not be available in many countries, making outreach efforts the primary way to engage such young people. In a response to the IAYC survey$^1$, one 1982 participant remarked (reproduced from \citealt{2019NatAs...3.1043D}):

\begin{quote}I am one of eight children and my father died when I was 9 years old. None of my siblings went to college from school. My trip to the IAYC, sponsored by a national scientific organisation, confirmed and inspired my love of astrophysics, making me determined to pursue it as a career. I subsequently completed an undergraduate degree, Masters and a PhD in mathematics, with a specialty in general relativity.\end{quote}

For the IAYC to become a truly inclusive activity, IWA must consider ways to avoid the cycle of excluding participants from low GNI countries. IWA's current grant program (funded by donations, not camp fees) is insufficient to close this gap, nor does IWA have enough sponsorship to lower the camp fee. In the past, IWA has cooperated with organizations like the \textit{National Youth Development Trust}, which supported African participants to attend the IAYC in the late-2000s. Some participants have had success finding individual sponsorship from their own universities, governments, or local charitable organizations. Overall, these efforts have been limited, providing solutions which are short-term and with little impact on the overall future of the camp. IWA is run entirely by volunteers, therefore members have minimal time to devote to these aspects in addition to the organization of the IAYC itself.

If IWA were to fall into the feedback loop of staying in expensive, English-speaking host countries with participants able to afford higher fees, this would not be a threat to the future of the IAYC --- the camp is currently oversubscribed by at least a factor of two. However, that is not the future that IWA wants. IWA is exploring solutions such as inviting participants to ``pay it forward'' when they pay their fee (rather than relying solely on alumni donations), prioritizing participants from low GNI countries when considering grant applications, and working with previous IWA members to find sponsorship in low GNI countries. As an international organization pursuing astronomy education outreach, it is IWA's responsibility to do the hard work of making this outreach available to all. 


\acknowledgments \software{Matplotlib \citep{Hunter:2007}}

\bibliographystyle{aasjournal}
\bibliography{rnaasbib}{}  

\end{document}